\documentclass[9pt,twocolumn,twoside]{opticajnl}
\journal{josab}

\setboolean{shortarticle}{false}

\usepackage{balance}

\usepackage{tikz,xcolor,hyperref}
\definecolor{lime}{HTML}{A6CE39}  
\DeclareRobustCommand{\orcidicon}{%
	\begin{tikzpicture}
	\draw[lime, fill=lime] (0,0) 
	circle [radius=0.16] 
	node[white] {{\fontfamily{qag}\selectfont \tiny ID}};
	\draw[white, fill=white] (-0.0625,0.095) 
	circle [radius=0.007];
	\end{tikzpicture}
	\hspace{-2mm}
}
\foreach \x in {A, ..., Z}{%
	\expandafter\xdef\csname orcid\x\endcsname{\noexpand\href{https://orcid.org/\csname orcidauthor\x\endcsname}{\noexpand\orcidicon}}
}

\usepackage{siunitx}

\renewcommand{\eqref}[1]{\hyperref[{#1}]{\textup{Eq.~(\ref*{#1}})}}
\newcommand{\secref}[1]{\hyperref[{#1}]{\textup{Sec.~\ref*{#1}}}}
\newcommand{\tabref}[1]{\hyperref[{#1}]{\textup{Table~\ref*{#1}}}}
\newcommand{\appref}[1]{\hyperref[{#1}]{\textup{Appendix~\ref*{#1}}}}

\newcommand{\vect}[1]{\boldsymbol{#1}}  
\newcommand{\uvect}[1]{\hat{\vect{#1}}}    

\makeatletter
\newcommand\figref[1]{%
    \@ifnextchar\bgroup{\figref@double{#1}}{\figref@single{#1}}%
}
\newcommand\figref@single[1]{\hyperref[{#1}]{\textup{Fig.~\ref*{#1}}}~}
\newcommand\figref@double[2]{\hyperref[{#1}]{\textup{Fig.~\ref*{#1}}\textcolor{blue}{\textbf{#2}}}}
\makeatother

\allowdisplaybreaks

\title{Focal control and light tracing on curved surfaces with isotropic transformation medium}

\author[1,\dag]{Xiaoyu Zhao\orcidA}
\author[1,\dag]{Longfei Shi\orcidB}
\author[1]{Zhuoyu Zhang\orcidC}
\author[1]{Xiaoke Gao\orcidD}
\author[1]{Jiawei Wang\orcidE}
\author[1]{\\Xikui Ma}
\author[1,*]{Tianyu Dong\orcidF} 

\affil{School of Electrical Engineering, Xi'an Jiaotong University, Xi'an 710049, China}
\affil[$\dag$]{These authors contribute equally.}
\affil[*]{Corresponding author: tydong@mail.xjtu.edu.cn}

\begin{abstract}
    Optics related to non-Euclidean geometry has been attracting growing interest for emerged novel phenomena and the analog for general relativity, while most studies are limited to the free space on rotationally-symmetric surfaces. In this paper, we focus on the light control and ray tracing on complex surfaces filled with inhomogeneous transformation medium. Within the conformal transformation optics, focal control devices and absolute optical instruments have been extended to curved surfaces. According to the equivalence between geometry and material, the metric tensor of the curved surface and the refractive index tensor are unified as the optical metric for the Hamilton's equations of light propagation on a curved surface. By solving for ray trajectories in the local coordinate system of mesh element and illuminating the refraction between non-planar elements with discontinuous media, a mesh-based ray-tracing algorithm on curved surface with medium has been proposed to validate the light control. Our research establishes a theoretical framework for light ray control in non-Euclidean space and offers an efficient tool for ray tracing in inhomogeneous medium on curved surface.
\end{abstract}

\setboolean{displaycopyright}{false} 
\doi{} 

\begin{document}

\maketitle

\section{Introduction}
Electromagnetism in curved geometry is important in the field of general relativity. Considering the difficulty in investigating non-Euclidean spacetime on a cosmic scale, researchers turn to the electromagnetic analogy in low-dimensional curved space, such as the simulation of the gravitational effect of a black hole \cite{leonhardt2002laboratory,philbin2008fiber,xu2021theory}. The motion of charged particles restricted on a surface influenced by curvature potential is also an important topic in microscopic physics \cite{da1981quantum,gravesen2008electron}. In recent years, optical research has been focusing on the impact of curved metric on the light propagation behavior, such as the refocusing and self-imaging governed by constant positive Gaussian curvature \cite{schultheiss2010optics}, the chaotic phenomenon during light propagation affected by geometric parameters \cite{liang2020mimicking,xu2022light}, and the diffraction of light beams \cite{bvelin2021optical,zhang2024diffraction}. Meanwhile, the theory of transformation optics \cite{pendry2006controlling} allows for studying the curved metric in equivalent material \cite{wang2018wave}, which is consistent with the design of light control devices based on transformation media induced from curved geometry such as geodesic lenses \cite{righini1973geodesic,xu2019conformal,ge2024geodesic}.

The development of electromagnetism reveals the electromagnetic wave nature of light, suggesting that light can be controlled by inhomogeneous media \cite{kong2008Electromagnetic}, such as achieving perfect imaging by Maxwell's fisheye \cite{wang2017self}. Following the idea that geometry and material are equivalent for light, a series of absolute optical instruments (AOIs) with refractive index distribution equivalent to spherically symmetric geometry were invented, such as the Luneburg lens \cite{luneburg1964mathematical}, Eaton lens \cite{eaton1952on}, optical black hole \cite{narimanov2009optical} and invisible sphere \cite{hendi2006ambiguities,minano2006perfect}. Furthermore, AOIs can be extended on curved surfaces for manipulating electromagnetic surface waves \cite{liu2010transformational,mitchell2014lenses} and Lissajous lens  \cite{danner2015lissajous,tyc2015absolute}. The study of transformation optics and AOIs has laid the foundation for light control on curved surfaces. However, previous research is basically restricted on a surface with rotational symmetry \cite{thomas2013perfect,schultheiss2020light}. The light manipulation on complex curved surfaces has long been hindered by the difficulty of mapping construction and transformation distortion that will cause strong anisotropy in medium \cite{mcmanus2014illusions,mcmanus2016isotropic}. In recent, the so-called boundary first flattening method \cite{sawhney2017boundary} has been introduced to construct quasi-conformal mapping and induce isotropic transformation medium that achieves a convex lens on piecewise analytic surface with circle base, while the potential of conformal parameterization methods \cite{chen2022handbook} for light control has not been fully exploited \cite{xu2020conformal}. A mesh-based solution method for isotropic transformation medium is proposed, which realizes electromagnetic surface wave manipulation on non-rotationally symmetric surfaces, with less attention paid on the light trajectory control \cite{zhao2023controlling}. In addition, in contrast to the rotationally symmetric surface where the light propagation can be easily solved, efficient ray tracing solution tools on complex surfaces are necessary for validating the light control design. Moreover, research of radar creeping wave keeps exploring efficient algorithms for ray tracing on surface \cite{chen2013ray,fu2015creeping,fu2015new}, but actually calculating geodesic lines in non-Euclidean free space. 

In this paper, we propose a light ray control method on curved surfaces within the framework of conformal transformation optics. The mesh-based spherical conformal mapping (SCM) method \cite{choi2015flash} is introduced to construct quasi-conformal mapping between curved surfaces, from which the isotropic transformation medium can be induced \cite{zhao2023controlling} to guarantee the light propagation in physical space exhibits the behavior designed in virtual space. To overcome the drawback that the mapping target of the specific point is uncontrollable, a multiple conformal mapping cascaded by M\"{o}bius transformation, the stereographic and inverse stereographic projection is proposed to reposition a pair of source and focal points, such that the equivalent medium can produce a focal control device on surfaces. Under the premise of geometric optics approximation, Hamilton's equations with the optical metric that unifies the metric and medium are deduced and solved in the local coordinate system of the mesh element. The refraction on the element boundary caused by mesh non-planarity and medium discontinuity is discussed in detail. Finally, the AOIs and focal control device on curved surfaces are validated by the proposed ray-tracing method on a meshed surface filled with isotropic transformation medium.

\section{Theory and methods}
\subsection{Conformal transformation optics}
\begin{figure} [!ht]
    \centering
    \includegraphics[width = \linewidth]{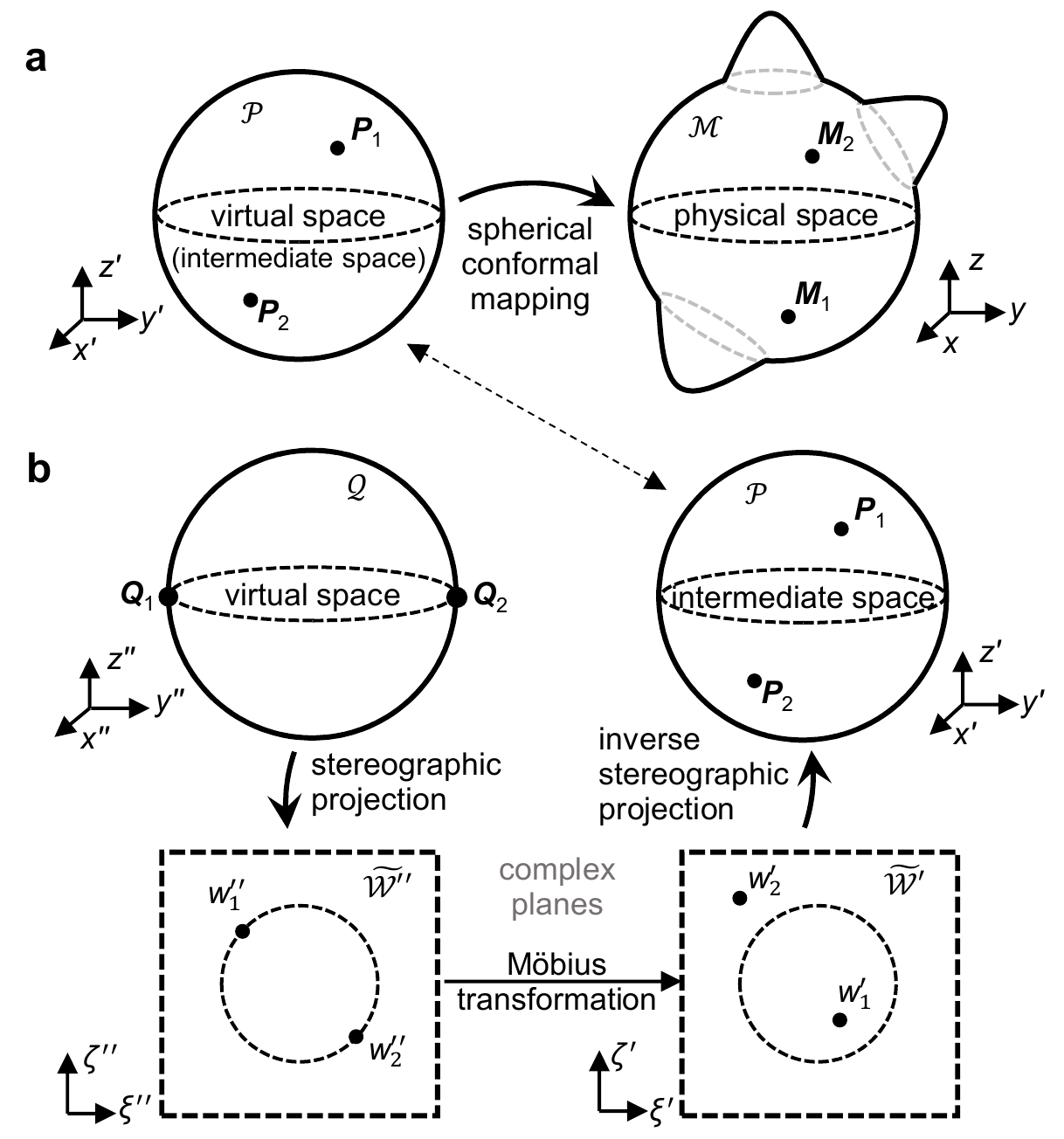}
    \caption{\textbf{Controlling light rays by virtue of mappings between curved surfaces}. (a) Quasi-conformal spherical conformal mapping (SCM) from a unit spherical surface $\mathcal{P}$ to a curved surface $\mathcal{M}$. (b) Multiple conformal mapping from a unit spherical surface $\mathcal{Q}$ to $\mathcal{P}$, repositioning locations of a pair of points (from points $\vect{Q}_1$ and $\vect{Q}_2$ to points $\vect{P}_1$ and $\vect{P}_2$, respectively), by cascading M\"{o}bius transformation, stereographic and inverse stereographic projections.}
    \label{fig:figure01}
\end{figure}

\figref{fig:figure01}{(a)} illustrates the light control on a curved surface $\mathcal{M}$ in the light of conformal transformation optics, which can be achieved by finding a conformal map between the arbitrary surface $\mathcal{M}$ and a unit spherical surface $\mathcal{P}$ with an isotropic transformation medium that can be deduced from the map $f: \mathcal{P} \rightarrow \mathcal{M}$. For instance, the so-called bijective spherical conformal mapping between the curved surface $\mathcal{M}$ and the unit spherical surface $\mathcal{P}$ can be used, which maps the points $\vect{P}_1$ and $\vect{P}_2$ in virtual space $\mathcal{P}$ to the points $\vect{M}_1$ and $\vect{M}_2$ in physical space $\mathcal{M}$, respectively. Therefore, the transformation medium $\vect{n}_\text{trans}$ induced from the mapping $f$ can yield electromagnetic phenomena on the physical curved surface $\mathcal{M}$ identical to those observed on the virtual spherical surface $\mathcal{P}$. 

Within the Einstein summation notation throughout, we denote  the surface $\mathcal{M}$ being analytically parameterized as $x^i = x^i(x^\alpha)$ with coordinates $x^\alpha = \{u, v\}$ and the virtual space $\mathcal{P}$ being also parameterized as $x^{i'} = x^{i'}(x^{\alpha'})$ with $x^{\alpha} = \{u', v'\}$, and the map $f$ is denoted by $x^i = x^i(x^{i'})$. Thus, the transformation medium on $\mathcal{M}$ that is induced from the map $f$ can be expressed as
\begin{equation} \label{eq:transformation-medium}
    \epsilon^{\alpha\beta} = \Lambda^{\alpha}_{\alpha'} \Lambda^{\beta}_{\beta'} \epsilon^{\alpha'\beta'},
\end{equation}
where $\epsilon = n^2$ with $n$ being the refractive index, $\epsilon^{\alpha'\beta'} = \delta^{\alpha'\beta'}$ when the virtual space $\mathcal{P}$ is a vacuum (here the Kronecker symbol $\delta^{\alpha'\beta'} = 1$ when $\alpha' = \beta'$; otherwise, $\delta^{\alpha'\beta'} = 0$). In addition, the Jacobian matrix $\Lambda^\alpha_{\alpha'} = \partial x^\alpha / \partial x^{\alpha'}$ between the parameterized spaces can be expressed as $\Lambda^{\alpha}_{\alpha'} = \Lambda^{\alpha}_{i} \Lambda^{i}_{i'} \Lambda^{i'}_{\alpha'}$, where the Jacobi matrices $\Lambda^{\alpha}_{i}$, $\Lambda^{i}_{i'}$ and $\Lambda^{i'}_{\alpha'}$ can be derived from the maps $x^i = x^i(x^\alpha)$, $x^i = x^i(x^i)$ and $x^i = x^i(x^\alpha)$, respectively. Alternatively, $\Lambda^{\alpha}_{\alpha'}$ may be determined by the metric tensors $g_{\alpha\beta}$ since $g_{\alpha'\beta'} = \Lambda^{\alpha}_{\alpha'} \Lambda^{\beta}_{\beta'} g_{\alpha\beta}$, where $g_{\alpha\beta} = \vect{e}_\alpha \cdot \vect{e}_\beta$ with $\vect{e}_\alpha = \partial{\vect{r}}/\partial x^{\alpha}$ and $\vect{r}$ being the position vector. When the conformality of the map $x^{i} = x^{i}(x^{i'})$ (or equivalently, $x^{\alpha} = x^{\alpha}(x^{\alpha'})$ is high enough, i.e., the conformal ratio $\mathcal{Q} = \text{max}(\sigma_{\vect{\Lambda}1} / \sigma_{\vect{\Lambda}2}, \sigma_{\vect{\Lambda}2} / \sigma_{\vect{\Lambda}1}) < 1.05$, the transformation medium \eqref{eq:transformation-medium} can be expressed as \cite{horman2008mesh}
\begin{equation} \label{eq:refractive-jacobi-metric}
    n_\text{trans} = 1/\sqrt{\sigma_{\vect{\Lambda}1} \sigma_{\vect{\Lambda}2}} = \sqrt{\sigma_1 \sigma_2 / (\sigma'_1 \sigma'_2)},
\end{equation}
where $\sigma_{\vect{\Lambda}1}$ and $\sigma_{\vect{\Lambda}2}$ are the singular values of the Jacobi matrix $\vect{\Lambda}$, also the scaling ratios of length element in two orthogonal directions; $\sigma'_1$ and $ \sigma'_2$ are singular values of the metric tensor $g_{\alpha'\beta'}$; $\sigma_1$ and $\sigma_2$ are the singular values of $g_{\alpha\beta}$, respectively. When the map $f$ is  conformal, we have
$n_\text{trans}(x^i) = 1/\sigma_{\vect{\Lambda}}(x^i)$, where $\sigma_{\vect{\Lambda}} = \sigma_{\vect{\Lambda}1} = \sigma_{\vect{\Lambda}2} = [\sigma'_1 \sigma'_2 / (\sigma_1 \sigma_2)]^{1/4}$. Consequently, the optical metric tensor ${h}_{\alpha\beta} = n_\text{trans}^2 {g}_{\alpha\beta}$ on $\mathcal{M}$ is equivalent to the metric tensor ${g}_{\alpha'\beta'}$ on $\mathcal{P}$, and light on $\mathcal{M}$ will exhibit the same propagation behavior as that in virtual space.

\subsection{Media of AOIs and focal control device on curved surface}
Often, we can begin with an absolution optical instrument in the virtual space $\mathcal{P}$ (unit spherical surface), which is not empty and filled with a refractive index distribution $n'_\text{AOIs}(x^{i'})$. As a result, the refractive index of AOIs on the curved surface $\mathcal{M}$ is the superposition of the AOI's medium $n'_\text{AOIs}$ in virtual space $\mathcal{P}$ and the transformation medium $n_\text{trans}$, which reads
\begin{equation} \label{eq:transformation-medium-physical-surfaces-AOIs}
    n_\text{AOIs}(x^i) = n'_\text{AOIs}(x^{i}) \cdot n_\text{trans}(x^i), 
\end{equation}
where $n'_\text{AOIs}(x^{i}) = n'_\text{AOIs}(x^{i'}(x^i))$. The light propagation behavior in the medium $n_\text{AOIs}$ filled on the curved surface $\mathcal{M}$ will be identical to that in the medium $n'_\text{AOIs}$ on the spherical surface $\mathcal{P}$. 

While optimizing the conformality of mapping between surfaces, most surface parameterization methods, such as SCM, cannot control the target position of specific points as desired. By cascading conformal mappings, a pair of points on the spherical surface can be repositioned, as shown in \figref{fig:figure01}{(b)}. Now, the unit spherical surface $\mathcal{P}$ in \figref{fig:figure01}{(a)} is regarded as an intermediate space between the virtual space $\mathcal{Q}$ and the physical space $\mathcal{M}$. With the stereographic projection $\xi'' = x'' / (1 - z'')$ and $\zeta'' = y'' / (1 - z'')$, points $\vect{Q}_1 = (x''_1, y''_1, z''_1)$ and $\vect{Q}_2 = (x''_2, y''_2, z''_2)$ on the unit spherical surface $\mathcal{Q}$ are mapped to $w''_1 = \xi''_1 + \text{i} \zeta''_1$ and $w''_2 = \xi''_2 + \text{i} \zeta''_2$ on the complex plane $\widetilde{\mathcal{W}}''$, respectively; within the inverse stereographic projection $x' = 2\xi' / (1 + \xi'^2 + \zeta'^2)$, $y' = 2\zeta' / (1 + \xi'^2 + \zeta'^2)$ and $z' = (\xi'^2 + \zeta'^2 - 1) / (1 + \xi'^2 + \zeta'^2)$, points $\vect{P}_1 = (x'_1, y'_1, z'_1)$ and $\vect{P}_2 = (x'_2, y'_2, z'_2)$ on the spherical surface $\mathcal{P}$ (intermediate space) are related to $w'_1 = \xi'_1 + \text{i} \zeta'_1$ and $w'_2 = \xi'_2 + \text{i} \zeta'_2$ on the complex plane $\widetilde{\mathcal{W}}'$, respectively. In addition, the points $w''_1$ and $w''_2$ on the complex plane $\widetilde{\mathcal{W}}''$ and $w'_1$ and $w'_2$ on the complex plane $\widetilde{\mathcal{W}}'$ are respectively corresponded by virtue of the M\"{o}bius transform, which reads
\begin{equation} \label{eq:mobius-transformation-formula}
    w' = w'(w'') = \dfrac{w'' + b}{w'' + d},
\end{equation}
where the parameters $b = [w'_1 w_2 (1 - w_1) + w_1 w'_2 (w_2 - 1)] / (w_1 - w_2)$ and $d = [w'_1 (1 - w_1) + w'_2 (w_2 - 1)] / (w_1 - w_2)$ should satisfy $d - b = [(w'_1 - 1)(w'_2 - 1)(w''_1 - w''_2)] / (w'_1 - w'_2)$, implying that $w'_1 \neq w'_2$, $w''_1 \neq w''_2$, $w'_1 \neq 1$ and $w'_2 \neq 1$. The theory of conformal transformation optics \cite{zhao2023controlling} has illuminated that the isotropic transformation medium of multiple conformal mappings can be formulated as the superposition of the medium induced from each mapping. Therefore, the transformation in the intermediate space $\mathcal{P}$ can be expressed as
\begin{equation} \label{eq:Mobius-transformation-medium}
    n_\text{inter} = n_\text{stereo} \cdot n_\text{M\"{o}bius} \cdot n_\text{invStereo},
\end{equation}
where $n_\text{stereo} = \sigma_{\vect{\Lambda}_\text{invStereo}} = 2 / (\xi''^2 + \zeta''^2 + 1)$, $n_\text{M\"{o}bius} = \sigma^{-1}_{\vect{\Lambda}_\text{M\"{o}bius}} = |\text{d}w''/\text{d}w'|$ and $n_\text{invStereo} = \sigma^{-1}_{\vect{\Lambda}_\text{invStereo}} = (\xi'^2 + \zeta'^2 + 1) / 2$ are the refractive indices of equivalent media of the stereographic projection, the M\"{o}bius transformation \cite{ulf2006optical} and the inverse stereographic projection, respectively. Note that the reciprocal relation between $n_\text{stereo}$ and $n_\text{invStereo}$ originates from the pseudo-inverse relation between the Jacobi matrices $\vect{\Lambda}_\text{stereo}$ and $\vect{\Lambda}_\text{invStereo}$ \cite{horman2008mesh}. If the points $\vect{Q}_1$ and $\vect{Q}_2$ are symmetrical about the sphere center in the virtual space $\mathcal{Q}$, light rays emitted from one point will always pass through the other point on the free-space spherical surface \cite{xu2019conformal}; therefore, the medium $n_\text{inter}$ will serve as a focal control device on the intermediate spherical surface $\mathcal{P}$ on which light rays emitted from the source $\vect{P}_1$ will focus at $\vect{P}_2$. Furthermore, a focal control device can be realized on the arbitrary closed surface $\mathcal{M}$ by incorporating the transformation medium $n_\text{trans}$ (see \eqref{eq:refractive-jacobi-metric}) induced from the quasi-conformal mapping between the intermediate space $\mathcal{P}$ and the physical space $\mathcal{M}$ (see \figref{fig:figure01}{(a)}), whose refractive index reads
\begin{equation} \label{eq:transformation-medium-physical-surfaces-focal}
    n_\text{Focal}(x^i) = n_\text{inter}(x^i) \cdot n_\text{trans}(x^i),
\end{equation}
where $n_\text{inter}(x^{i}) = n_\text{inter}(x^{i'}(x^i))$; and the points $\vect{M}_1$ and $\vect{M}_2$ are the source and focus of the focal control device, respectively.

\subsection{Hamilton's equations on curved surfaces filled with medium}
Ray tracing on a curved surface relies on solving Hamilton's equations with metric. Suppose a space $\mathcal{M}$ described by a coordinate system $\{x^\gamma\}$ with a metric tensor $\vect{g} = g_{\alpha\beta}(x^\gamma)$ defining the optical path length element $\text{d} \ell^2 = g_{\mu\nu} \text{d} x^\mu \text{d} x^\nu$ in free space, with $\text{d}x^\mu$ being the length element in geometry. If the space is filled with an inhomogeneous anisotropic medium, the optical path length element can be rewritten as $\text{d}\ell^2 = g_{\alpha\beta} n^\alpha_\mu n^\beta_\nu \text{d} x^\mu \text{d} x^\nu = h_{\mu\nu} \text{d} x^\mu \text{d} x^\nu$, where $h_{\mu\nu}(x^\gamma) = g_{\alpha\beta} n^\alpha_\mu n^\beta_\nu$ is the optical metric tensor for a space $\mathcal{M}$ with metric tensor $g_{\alpha\beta}(x^\gamma)$ and medium $\vect{n} = n^\alpha_\mu (x^\gamma)$. From the equivalence between geometry and medium, the refractive index $n^\alpha_\mu$ acts on the length element $\text{d} x^\mu$ like the Jacobi matrix $\Lambda^\alpha_\mu = \partial x^\alpha / \partial x^\mu$, and $n^\alpha_\mu n^\beta_\nu$ is equivalent to endowing the space $\mathcal{M}$ with a new metric. Therefore, the light propagation in the space $\mathcal{M}$ can be regarded as the propagation in free space with a new metric tensor $\vect{h}$, and Hamilton's equations read as \cite{leonhardt2010geometry}: 
\begin{subequations} \label{eq:hamilton-new-metric}
    \begin{align}
        \dfrac{\text{d} x^\alpha}{\text{d}t} &= \dfrac{\partial \omega}{\partial k_\alpha} = \dfrac{c_0 h^{\alpha\beta} k_\beta}{\sqrt{h^{\mu\nu} k_\mu k_\nu}}, \\
        \dfrac{\text{d} k_\alpha}{\text{d} t} &= - \dfrac{\partial \omega}{\partial x^\alpha} = - \dfrac{c_0 k_\kappa k_\eta }{2\sqrt{h^{\mu\nu} k_\mu k_\nu}} \dfrac{\partial h^{\kappa\eta}}{\partial x^\alpha}, 
    \end{align}
\end{subequations}
where $t$ is the time, $\omega = c \sqrt{h^{\mu\nu} k_\mu k_\nu}$ is the frequency, and $c_0$ is the light velocity in vacuum. Here, $k_\alpha$ denotes the phase gradient $k_\alpha = \partial_\alpha \varphi = g_{\alpha\beta} k^\beta$, where $k^\alpha$ denotes the component of the propagation vector $\vect{k} = k^\alpha \vect{e}_\alpha$. Note that \eqref{eq:hamilton-new-metric} is deduced from electromagnetic wave equations under the approximation of geometrical optics, which requires that the wavelength $\lambda = 2\pi / k$ (with $k = |\partial_\alpha\varphi|$ being the wave number) varies little over short distances ($|\nabla \lambda| \ll 1$) and the curvature is small on the scale of wavelength ($|R^2_{\alpha\beta}| \lambda^2 \ll 1$; $R_{\alpha\beta}$ is the Ricci curvature tensor of the curved geometry $\mathcal{M}$)\cite{leonhardt2010geometry}.  For an isotropic medium $\vect{n} = n \delta^\alpha_\beta(x^\gamma)$, the optical metric tensor reads $h_{\mu\nu} = g_{\alpha\beta} \delta^\alpha_\mu \delta^\beta_\nu n^2 = g_{\mu\nu} n^2$ with the inverse optical metric tensor being $h^{\mu\nu} = g^{\mu\nu}/n^2$; consequently, the Hamilton's equations \eqref{eq:hamilton-new-metric} can be simplified as
\begin{subequations} \label{eq:hamilton-isotropic-medium}
    \begin{align}
        \dfrac{\text{d} x^\alpha}{\text{d}t} &= \dfrac{c_0 k_\beta g^{\alpha\beta}/n^2}{ \sqrt{k_\mu k_\nu g^{\mu\nu}/n^2}} = \dfrac{c_0 g^{\alpha\beta} k_\beta}{n \sqrt{k_\mu k_\nu g^{\mu\nu}}}, \label{eq:hamilton-isotropic-medium-1} \\
        \dfrac{\text{d} k_\alpha}{\text{d} t} &= - \dfrac{c_0 k_\kappa k_\eta}{2 \sqrt{k_\mu k_\nu g^{\mu\nu}/n^2}} \dfrac{\partial \left( g^{\kappa\eta}/n^2 \right)}{\partial x^\alpha} \notag \\
        &= - \dfrac{c_0 k_\kappa k_\eta}{2n\sqrt{g^{\mu\nu} k_\mu k_\nu}} \left( \dfrac{\partial g^{\kappa\eta}}{\partial x^\alpha} - 2 \dfrac{g^{\kappa\eta}}{n} \dfrac{\partial n}{\partial x^\alpha} \right).  \label{eq:hamilton-isotropic-medium-2} 
    \end{align}
\end{subequations}
Given the metric tensor $g_{\alpha\beta}(x^\gamma)$ and the medium distribution $n(x^\gamma)$ on the curved surface $\mathcal{M}$, and the initial conditions $x^\alpha_0 = x^\alpha(t = 0)$ and $k_{\alpha0} = k_\alpha(t = 0)$, the Hamilton's equations \eqref{eq:hamilton-isotropic-medium} can be readily solved by numerical techniques such as the Runge--Kutta method.

\subsection{Local coordinate system on mesh element}
\begin{figure}[!ht]
    \centering
    \includegraphics[width = \linewidth]{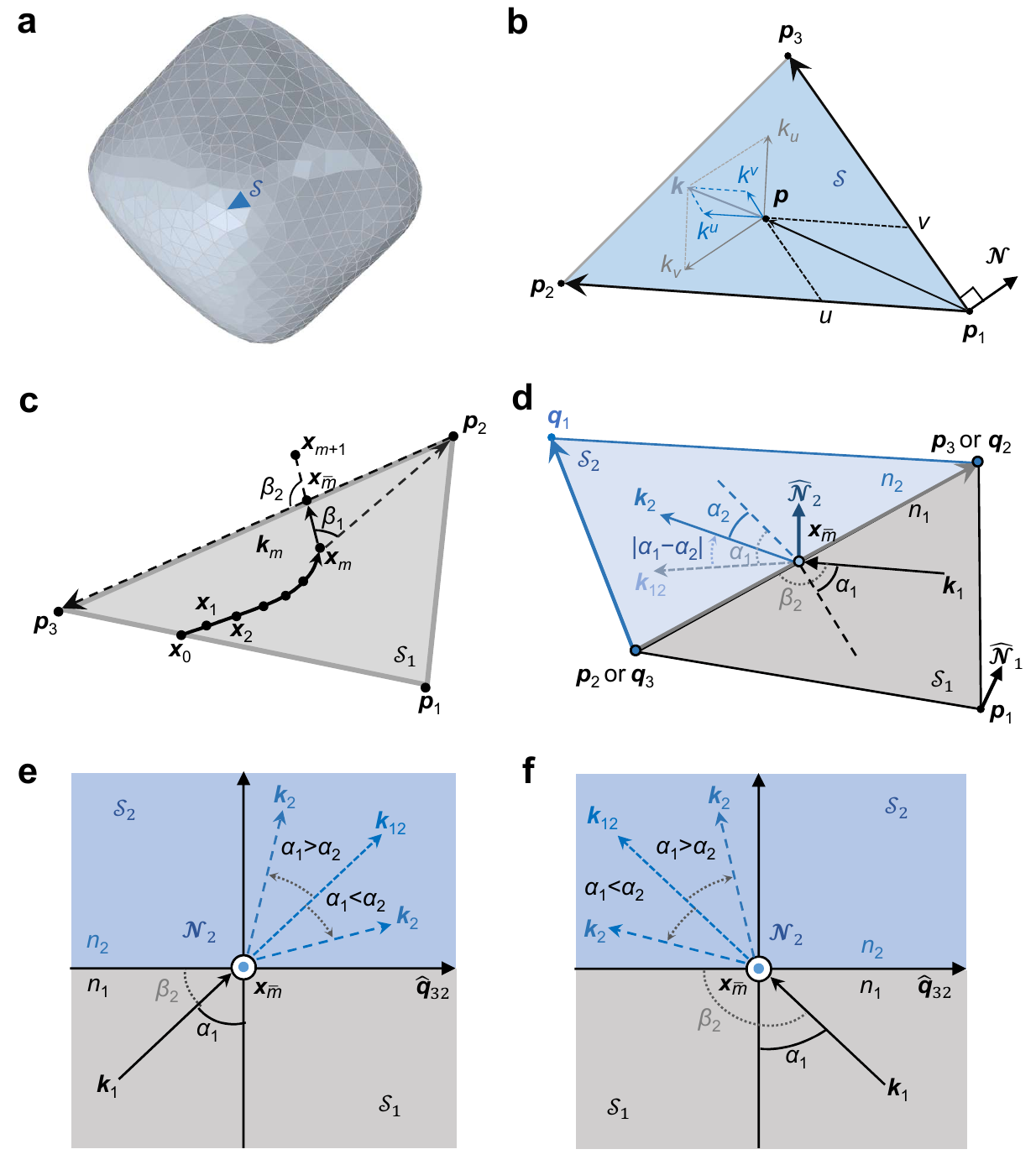}
    \caption{\textbf{Ray tracing on discretized surfaces}. (a) Triangulated representation of  $\mathcal{M}_\text{octa}$ with the blue region $\mathcal{S}$ denoting a triangle mesh element. (b) The local coordinate system on the triangulated mesh element $\mathcal{S}$. (c) Location of the boundary point. (d) Refraction of light ray between two meshes. Rotation of projection vector $\vect{k}_{12}$ when (e) $\beta_2 < {\pi}/{2}$ and (f) $\beta_2 > {\pi}/{2}$.}
    \label{fig:figure02}
\end{figure}

In fact, most curved surfaces lack global analytical parameterization, which provides stages for conformal parameterization methods based on surface meshing \cite{choi2015flash, chen2022handbook}. For example, when the surface is discretized by triangle patches, vertices and mesh elements are described by an $N_V \times 3$ vertex array $\vect{V}$ and an $N_F \times 3$ mesh array $\vect{F}$, respectively, as shown by the discretized octahedral
surface $\mathcal{M}_\text{octa}$ in \figref{fig:figure02}{(a)}. Now, the map $x^\alpha = x^\alpha(x^{\alpha'})$ can be handled in the local coordinate system $\{x^\alpha\}$ and $\{x^{\alpha'}\}$ in each triangulated mesh element. Therefore, the ray tracing equations \eqref{eq:hamilton-new-metric} need to be tailored for the local coordinates $x^\alpha$, including the position, wave vector, metric, and refractive index.

Taking triangular mesh for example, any point $\vect{p} = (x, y, z)^\text{T}$ inside a triangular element can be represented by its vertices $\vect{p}_1 = (x_1, y_1, z_1)^\text{T}$, $\vect{p}_2 = (x_2, y_2, z_2)^\text{T}$, and $\vect{p}_3 = (x_3, y_3, z_3)^\text{T}$ as $\vect{p} = \vect{p}_1 + u \vect{e}_u + v \vect{e}_v$, where $u = S_{\triangle\vect{p}_1\vect{p}\vect{p}_3}/S_{\triangle\vect{p}_1\vect{p}_2\vect{p}_3}$ and $v = S_{\triangle\vect{p}_1\vect{p}\vect{p}_2}/S_{\triangle\vect{p}_1\vect{p}_2\vect{p}_3}$ are the barycentric coordinates ($S$ denotes the area of the triangle), as shown in \figref{fig:figure02}{(b)}; $\vect{e}_u = \vect{p}_2 - \vect{p}_1 $ and $ \vect{e}_v = \vect{p}_3 - \vect{p}_1$ are the basis vectors with $\vect{p}_1$ denoting the origin of the local coordinate system $x^{\alpha} = \{u, v\}$. The transformations between the global Cartesian coordinates $x^i$ and local coordinates $x^{\alpha}$ can be expressed in a more compact form as:
\begin{subequations} \label{eq:uv-xyz-tensor}
    \begin{align}
        x^i &= p^i_1 + \Lambda^i_{\alpha, \mathcal{S}}x^\alpha_{\mathcal{S}}, \label{eq:uv-2-xyz-tensor} \\
        x^\alpha_{\mathcal{S}} &= J^\alpha_{i, \mathcal{S}} x^i, \label{eq:xyz-2-uv-tensor}
    \end{align}
\end{subequations}
where $\Lambda^i_{\alpha, \mathcal{S}} = \partial x^i / \partial x^\alpha_{\mathcal{S}}$ and $J^\alpha_{i, \mathcal{S}} = \partial x^\alpha_{\mathcal{S}} / \partial x^i$ denote the Jacobi matrices of \eqref{eq:uv-2-xyz-tensor} and \eqref{eq:xyz-2-uv-tensor}, respectively. The relationship between the phase gradient $k_{\alpha, \mathcal{S}} $ in the local coordinate system and the propagation vector $k^i $ in the global Cartesian system can be denoted as
\begin{subequations} \label{eq:wave-vector-transformation}
\begin{align}
    k_{\alpha, \mathcal{S}} &= g_{\alpha\beta, \mathcal{S}} J^\beta_{i, \mathcal{S}} k^i, \label{eq:wave-vector-transformation-1} \\
    k^i &= \Lambda^i_{\beta, \mathcal{S}} g^{\alpha\beta}_{\mathcal{S}} k_{\alpha,\mathcal{S}}. \label{eq:wave-vector-transformation-2}
\end{align}
\end{subequations}
where $g^{\alpha\beta}_{\mathcal{S}} = ( g_{\alpha\beta, \mathcal{S}} )^{-1}$ is the inverse metric tensor. Details of the transformations of \eqref{eq:uv-xyz-tensor} and \eqref{eq:wave-vector-transformation} are provided in \appref{app:matrix-form-transformation}. In addition, since the metric tensor $g_{\alpha\beta, \mathcal{S}} = \vect{e}_\alpha \cdot \vect{e}_\beta$ is a constant over $\mathcal{S}$, the derivative of $g_{\alpha\beta, \mathcal{S}}$ with respect to $x^\alpha$ vanishes in \eqref{eq:hamilton-isotropic-medium}. 

According to \eqref{eq:refractive-jacobi-metric}, the isotropic refractive index of the transformation medium can be induced from the mapping between the mesh elements $\mathcal{S}'$ and $\mathcal{S}$, which reads as
\begin{equation}
    n_\text{trans}(F_{\mathcal{S}}) = \sqrt{\sigma_{\mathcal{S}1} \sigma_{\mathcal{S}2} / (\sigma'_{\mathcal{S'}1} \sigma'_{\mathcal{S'}2})},
\end{equation}
where $\sigma_{\mathcal{S}1}$ and $\sigma_{\mathcal{S}2}$ are singular values of the metric tensor $g_{\alpha\beta, \mathcal{S}}$ in the simplex $\mathcal{S}$ on the discretized surface $\mathcal{M}$; $\sigma'_{\mathcal{S'}1}$ and $\sigma'_{\mathcal{S'}2}$ are singular values of the metric tensor $g_{\alpha'\beta', \mathcal{S'}}$ in the simplex $\mathcal{S}'$ on the discretized surface $\mathcal{P}$. In addition, $F_\mathcal{S}$ is the index of the mesh element $\mathcal{S}$, ranging from 1 to $N_F$. Therefore, the functional media for AOIs \eqref{eq:transformation-medium-physical-surfaces-AOIs} and focal control device \eqref{eq:transformation-medium-physical-surfaces-focal} can be respectively rewritten as 
\begin{subequations} \label{eq:transformation-medium-on-physical-surfaces}
    \begin{align}
        n_\text{AOIs}(x^\alpha_{\mathcal{S}}) &= n'_\text{AOIs}(x^\alpha_{\mathcal{S}}) \cdot n_\text{trans}(F_{\mathcal{S}}), \label{eq:transformation-medium-on-physical-surfaces-1} \\
        n_\text{Focal}(x^\alpha_{\mathcal{S}}) &= n_\text{inter}(x^\alpha_{\mathcal{S}}) \cdot n_\text{trans}(F_{\mathcal{S}}), \label{eq:transformation-medium-on-physical-surfaces-2}
    \end{align}
\end{subequations}
where $x^\alpha_\mathcal{S} = \{u_{\mathcal{S}}, v_{\mathcal{S}}\}$ denotes the local coordinates of simplex $\mathcal{S}$; $n'_\text{AOIs}(x^\alpha_{\mathcal{S}}) = n'_\text{AOIs}(x^i(x^\alpha_{\mathcal{S}}))$ and $n_\text{inter}(x^\alpha_{\mathcal{S}}) = n_\text{inter}(x^i(x^\alpha_{\mathcal{S}}))$ can be determined according to \eqref{eq:uv-2-xyz-tensor}. Consequently, the derivative $\partial n / \partial x^\alpha$ in \eqref{eq:hamilton-isotropic-medium} can be easily calculated.

\subsection{Refraction between meshes} \label{sec:ray-tracing-method}
When employing numerical methods such as the Runge--Kutta method to solve \eqref{eq:hamilton-isotropic-medium} on a mesh, the ray tracing may cross the element boundary due to the finite step size. Hence, it is necessary to determine the intersection when the light trajectory reaches the boundary, which demands the criteria derived from the definition $\vect{p} = \vect{p}_1 + u \vect{e}_u + v \vect{e}_v$ to judge that the location $(u, v)^\text{T}$ is outside the triangle region $\triangle \vect{p}_1\vect{p}_2\vect{p}_3$: $u < 0$, $v < 0$ or $u+v > 1$. Suppose that the last solution inside the simplex $\mathcal{S}_1$ is $\vect{x}_m$ at step $m$, as illustrated in \figref{fig:figure02}{(c)}, the interaction point $\vect{x}_{\overline{m}}$ can be calculated through extending the light trajectory from $\vect{x}_m$ in the direction of wave vector $\vect{k}_{m}$, which reads
\begin{equation} \label{eq:location-boundary-point}
    \vect{x}_{\overline{m}} = \vect{p}_2 + |\vect{x}_{\overline{m}} - \vect{p}_2| \cdot \hat{\vect{p}}_{23} = \vect{p}_2 + |\vect{x}_m - \vect{p}_2| \cdot \dfrac{\sin{\beta_1}}{\sin{\beta_2}} \cdot \hat{\vect{p}}_{23},
\end{equation}
where $\uvect{p}_{23} = (\vect{p}_3 - \vect{p}_2) / |\vect{p}_3 - \vect{p}_2|$ denotes the unit vector pointing from $\vect{p}_2$ to $\vect{p}_3$; $\beta_1 = \arccos \left( \uvect{k}_m \cdot \uvect{p}_{m2} \right)$ and $\beta_2 = \arccos \left( \uvect{k}_m \cdot \uvect{p}_{23} \right)$ are determined by the law of cosines with $\uvect{k}_m = \vect{k}_m / |\vect{k}_m|$ and $\uvect{p}_{m2} = (\vect{p}_2 - \vect{x}_m) / |\vect{p}_2 - \vect{x}_m|$. Note that the barycentric coordinates of point $\vect{x}_m$, the boundary vertices $\vect{p}_2$ and $\vect{p}_3$ are sorted such that $|\vect{p}_2 - \vect{x}_{\overline{m}}| < |\vect{p}_3 - \vect{x}_{\overline{m}}|$ and $u + v \rightarrow 1$. The interaction point $\vect{x}_{\overline{m}}$, as the end position of ray tracing in simplex $\mathcal{S}_1$ (formed by $\vect{p}_1$, $\vect{p}_2$ and $\vect{p}_3$), is also the initial position of ray tracing in simplex $\mathcal{S}_2$ (formed by $\vect{q}_1$, $\vect{q}_2$ (or $\vect{p}_3$) and $\vect{q}_3$ (or $\vect{p}_2$)), as shown in \figref{fig:figure02}{(d)}. According to \eqref{eq:uv-2-xyz-tensor} and \eqref{eq:xyz-2-uv-tensor}, the transformation from local coordinates $x^\alpha_{\mathcal{S}_1}$ to $x^\alpha_{\mathcal{S}_2}$ reads as $x^\alpha_{\mathcal{S}_2} = J^\alpha_{i, \mathcal{S}_2} ( p^i_{1, \mathcal{S}_1} + \Lambda^i_{\beta, \mathcal{S}_1} x^\beta_{\mathcal{S}_1} )$, where $p^i_{1, \mathcal{S}_1}$ is the origin of the local coordinate system $x^\alpha_{\mathcal{S}_1} = \{u_{\mathcal{S}_1}, v_{\mathcal{S}_1}\}$ in simplex $\mathcal{S}_1$, $\Lambda^i_{\beta, \mathcal{S}_1}$ is the Jacobi matrix of the transformation from local coordinates $x^\beta_{\mathcal{S}_1}$ to global coordinates $x^i$ (see \eqref{eq:matrix-form-lambda}), and $J^\alpha_{i, \mathcal{S}_2}$ is the Jacobi matrix of the transformation from $x^i$ to $x^\alpha_{\mathcal{S}_2}$ (see \eqref{eq:matrix-form-J}). 

Now, we discuss the light refraction between the two simplexes $\mathcal{S}_1$ and $\mathcal{S}_2$. There are two refractions: the projective refraction since the $\mathcal{S}_1$  and $\mathcal{S}_2$ are not in-plane; the usual optical refraction due to the medium discontinuity between $n_{\mathcal{S}_1}(\vect{x}_{\overline{m}})$ and $n_{\mathcal{S}_2}(\vect{x}_{\overline{m}})$, as illustrated in \figref{fig:figure02}{(d)}. The projection of $\vect{k}_1$ in $\mathcal{S}_1$ onto $\mathcal{S}_2$ reads
\begin{equation} \label{eq:projection-wave-vector}
    \vect{k}_{12} = \uvect{\mathcal{N}}_2 \times \vect{k}_1 \times \uvect{\mathcal{N}}_2 = \vect{k}_1 - \uvect{\mathcal{N}}_2 (\vect{k}_1 \cdot \uvect{\mathcal{N}}_2),
\end{equation}
where $\uvect{\mathcal{N}}_2 = \uvect{q}_{32} \times \uvect{q}_{31}$ is the unit normal vector of the plane of $\mathcal{S}_2$; $\uvect{q}_{32} = (\vect{q}_2 - \vect{q}_3) / |\vect{q}_2 - \vect{q}_3|$ and $\uvect{q}_{31} = (\vect{q}_1 - \vect{q}_3) / |\vect{q}_1 - \vect{q}_3|$; $\vect{k}_1 = k^i_1\vect{e}_i$ with $k^i_1 = \Lambda^i_{\beta, \mathcal{S}_1} g^{\alpha\beta}_{\mathcal{S}_1} k_{\alpha, \mathcal{S}_1}$ being the component of the propagation vector in the global Cartesian coordinates system and $k_{\alpha, \mathcal{S}_1} = (k_{u, \mathcal{S}_1}, k_{v, \mathcal{S}_1})$ being the phase gradient in the local coordinate system $x^\alpha_{\mathcal{S}_1}$. Note that vertices $\vect{q}_2$ and $\vect{q}_3$ should maintain the positional relationship in \figref{fig:figure02}{(c)} for the convenience of subsequent vector calculation, i.e., $|\vect{q}_3 - \vect{x}_{\overline{m}}| < |\vect{q}_2 - \vect{x}_{\overline{m}}|$. For optical refraction, the refraction angle $\alpha_2$ in $\mathcal{S}_2$ is given by Snell's refraction law: $\alpha_2 = \arcsin [(n_1 / n_2) \sin{\alpha_1}]$, where $n_1 = n_{\mathcal{S}_1}(\vect{x}_{\overline{m}})$ and $n_2 = n_{\mathcal{S}_2}(\vect{x}_{\overline{m}})$; $\alpha_1 = |\pi/2 - \beta_2|$ is the incident angle in $\mathcal{S}_1$. Utilizing the Rodrigues formula for vector rotation \cite{dai2015euler}, the initial wave vector $\vect{k}_2$ in simplex $\mathcal{S}_2$ can be determined by rotating the projection vector $\vect{k}_{12}$ by an angle $\alpha_0$ around the normal vector $\vect{\mathcal{\hat{N}}}_2$, which reads as
\begin{equation} \label{eq:rodrigues-rotation-simple}
    \vect{k}_2 = \vect{k}_{12} \cos{\alpha_0} + (\vect{\mathcal{\hat{N}}}_2 \times \vect{k}_{12}) \sin{\alpha_0},
\end{equation}
where $\alpha_0 = \alpha_1 - \alpha_2$ when $\beta < \pi/2$ and $\alpha_0 = \alpha_2 - \alpha_1$ when $\beta > \pi/2$, obeying the right-handed rotation rule for the Rodrigues formula, as demonstrated in \figref{fig:figure02}{(e)} and \figref{fig:figure02}{(f)}. Note that the axis vector $\hat{\vect{p}}_{23}$ should maintain the relation $|\vect{p}_2 - \vect{x}_{\overline{m}}| < |\vect{p}_3 - \vect{x}_{\overline{m}}|$. Furthermore, when handling \eqref{eq:hamilton-isotropic-medium}, we need to transform $\vect{k}_2$ stated in the global coordinate system to the phase gradient $k_{\alpha, \mathcal{S}_2} = (k_{u, \mathcal{S}_2}, k_{v, \mathcal{S}_2})$ in the local coordinate system $\{x^\alpha_{\mathcal{S}_2}\}$, namely, $k_{\alpha, \mathcal{S}_2} = g_{\alpha\beta, \mathcal{S}_2} J^\beta_{i, \mathcal{S}_2} k^i$, where $g_{\alpha\beta, \mathcal{S}_2}$ is the metric tensor of the local coordinate system in simplex $\mathcal{S}_2$. The pseudo-code for ray tracing on a meshed surface filled with medium is shown in \appref{app:pseudocode-ray-tracing}.

\section{Results and Discussion}
\subsection{Absolute optical instruments on curved surface}
Within the proposed method, we can design absolute instruments on curved surfaces, e.g., optical black hole and Eaton lens. In the virtual space $\mathcal{P}$ denoted by a unit sphere, the refractive indices of the optical black hole and Eaton lens respectively read 
\begin{subequations} \label{eq:absolute-instrument-virtual}
    \begin{align}
        n'_\text{Black}(u'_{\mathcal{S'}}, v'_{\mathcal{S'}}) &= \dfrac{\pi}{\theta'(u'_{\mathcal{S'}}, v'_{\mathcal{S'}})}, \label{eq:absolute-instrument-virtual-1} \\
        n'_\text{Eaton}(u'_{\mathcal{S'}}, v'_{\mathcal{S'}}) &= \sqrt{\dfrac{2\pi}{\theta'(u'_\mathcal{S'}, v'_\mathcal{S'})} - 1}, \label{eq:absolute-instrument-virtual-2}
    \end{align}
\end{subequations}
where $\theta'(u'_{\mathcal{S}'}, v'_{\mathcal{S}'}) = \theta'(x', y', z') = \text{arctan} \left[\sqrt{x'^2(u'_{\mathcal{S}'}, v'_{\mathcal{S}'}) + y'^2(u'_{\mathcal{S}'}, v'_{\mathcal{S}'})} /z'(u'_{\mathcal{S}'}, v'_{\mathcal{S}'})  \right]$ is the polar angle of global spherical coordinates and $\{ x'(u'_{\mathcal{S}'}, v'_{\mathcal{S}'}), y'(u'_{\mathcal{S}'}, v'_{\mathcal{S}'}), z'(u'_{\mathcal{S}'}, v'_{\mathcal{S}'}) \}$ are the global Cartesian coordinates expressed by the local coordinates $\{u'_{\mathcal{S}'}, v'_{\mathcal{S'}}\}$ in the virtual space. Since the mapping $f: \mathcal{P} \rightarrow \mathcal{M}_\text{octa}$ is almost distortion-free with a high degree of conformality, the simplexes $\mathcal{S}'$ and $\mathcal{S}$ can share the same local coordinate system, and hence $u'_{\mathcal{S}'} = u_{\mathcal{S}}$ and $ v'_{\mathcal{S}'} = v_{\mathcal{S}}$ can lead to the transformation medium $n_\text{AOIs}(x^\alpha_{\mathcal{S}})$ given by \eqref{eq:transformation-medium-on-physical-surfaces-1}. The equivalence between geometry and medium implies a three-layer metric endowed on the curved surface $\mathcal{M}_\text{octa}$: the curved metric of the surface $\mathcal{M}_\text{octa}$ emulated by the Euclidean metric of mesh elements, the metric of the spherical surface $\mathcal{P}$ emulated by the transformation medium $n_\text{trans}$, and the equivalent metric of AOIs media $n'_\text{Black}$ or $n'_\text{Eaton}$.

\begin{figure}[!ht]
    \centering
    \includegraphics[width = \linewidth]{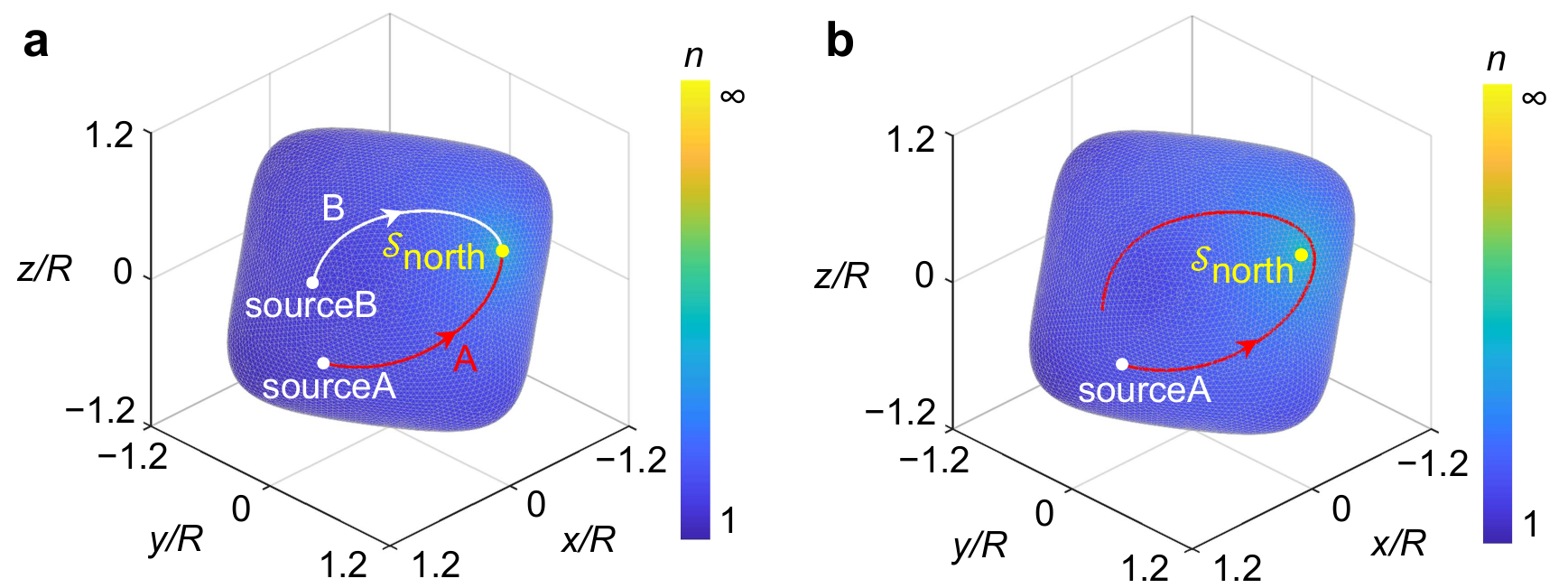}
    \caption{\textbf{Ray tracing in absolute optical instruments on curved surface $\mathcal{M}_\text{octa}$}. (a) Optical black hole. (b) Eaton lens. All dimensions are normalized to the radius $R = 1$ of the spherical surface in the virtual space $\mathcal{P}$. The light velocity $c_0 = 1$ is set to be dimensionless, and the propagation time in devices is $t_\text{EATON} = 12~R/c_0$ and $t_\text{BLACK} = 20~R/c_0$, respectively. The total number of mesh elements is $N_V = 18602$, the average length of mesh edges is $\overline{\Delta}_\text{mesh} = 0.05 R$ and the step size takes $\Delta t = 10^{-5}~R/c_0$. The device ``center'' is located at $\vect{r}_\text{north}= (-0.20, 0.89, 0.52)^\text{T}$ corresponding to the maximum refractive index. The locations of light sources are $\vect{r}_0^\text{A} = (1.00, 0.34, -0.12)^\text{T}$ and $\vect{r}_0^\text{B} = (1.05, 0.30, 0.56)^\text{T}$ in (a) and $\vect{r}_0 = \vect{r}_0^\text{A}$ in (b). The initial wave vectors are $\vect{k}_0^\text{A} = (-0.32, 0.95, 0)^\text{T}$ and $\vect{k}_0^\text{B} = (0.25, 0.08, 0.97)^\text{T}$ in (a) and $\vect{k}_0 = \vect{k}_0^\text{A}$ in (b). All the vectors are given in terms of the global Cartesian coordinates.}
    \label{fig:figure03}
\end{figure}

With the inverse metric tensor $g^{\alpha\beta}(F_{\mathcal{S}})$, the refractive index distribution $n_\text{trans}(F_{\mathcal{S}})$ of the transformation medium and AOIs media $n'_\text{Black}(u_{\mathcal{S}}, v_{\mathcal{S}})$ or $n'_\text{Eaton}(u_{\mathcal{S}}, v_{\mathcal{S}})$ on the triangulated mesh $\mathcal{M}_\text{octa}$, one can solve Hamilton's equations \eqref{eq:hamilton-isotropic-medium} in the local coordinate systems using the algorithm presented in \secref{sec:ray-tracing-method}. \figref{fig:figure03} displays the ray tracing in AOIs on the curved surface $\mathcal{M}_\text{octa}$. For the optical black hole presented in \figref{fig:figure03}{(a)}, light rays A and B propagate along an approximate spiral path and are eventually bounded at the device center $\mathcal{S}_\text{north}$ corresponding to the north pole of the unit sphere in the virtual space; and the light trajectory bypasses $\mathcal{S}_\text{north}$ and propagates in the opposite direction, behaving as a surface Eaton lens as shown in \figref{fig:figure03}{(b)}. The results of ray tracing validate the ability of the  optical black hole and Eaton lens to confine and bend light rays on the curved surface. Comparing \eqref{eq:absolute-instrument-virtual-1} and \eqref{eq:absolute-instrument-virtual-2}, though the refractive index distribution of the optical black hole and Eaton lens both reaches positive infinity at the device center, the function $n'_\text{Black}(\theta)$ increases faster than $n'_\text{Eaton}(\theta)$ when $\theta \rightarrow 0$. Consequently, the optical black hole has a stronger ability than the Eaton lens to confine light ray at $\mathcal{S}_\text{north}$.

\subsection{Focal control devices on curved surface}
\begin{figure}[!ht]
    \centering
    \includegraphics[width = \linewidth]{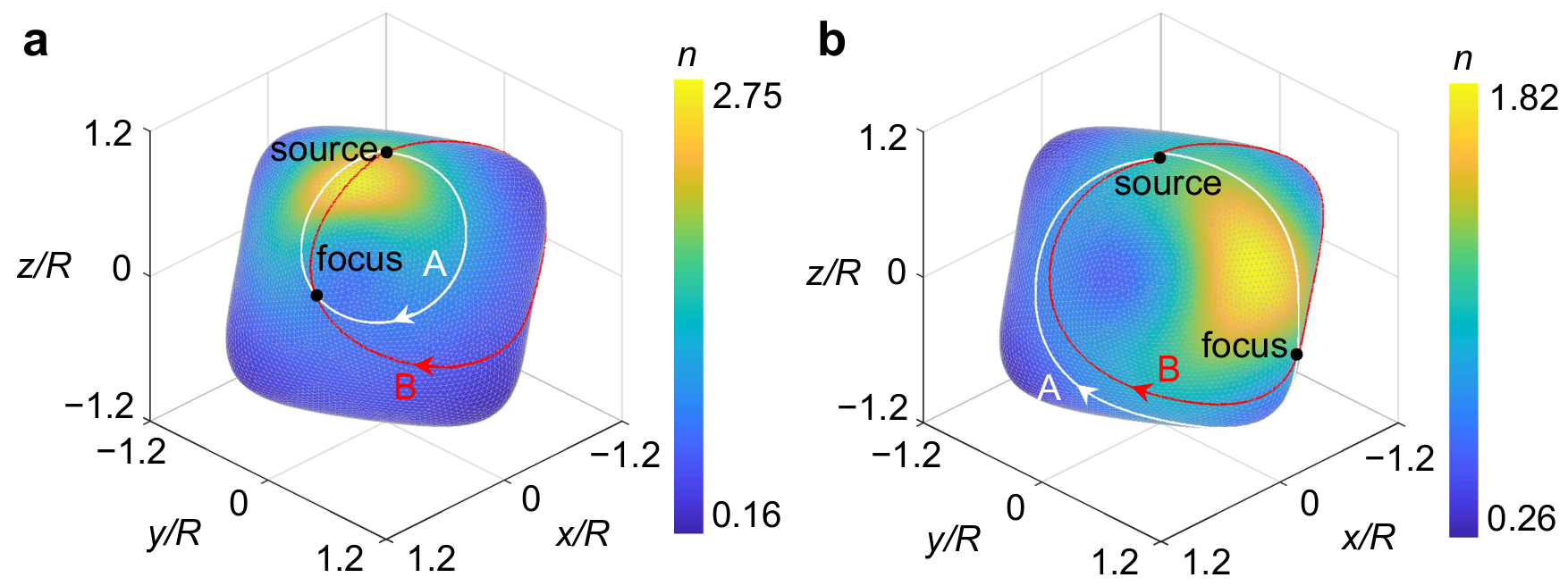}
    \caption{\textbf{Ray tracing in focal control devices on surface $\mathcal{M}_\text{octa}$ with different focal locations}: (a) $\vect{r}^\text{(a)}_\text{focal} = (1.08, 0.38, 0.44)^\text{T}$ and (b) $\vect{r}^\text{(b)}_\text{focal} = (-0.28, 1.10, -0.31)^\text{T}$. All coordinates are normalized to the radius $R$ of the spherical surface in the virtual space $\mathcal{P}$. The light velocity $c_0 = 1$ is set to be dimensionless, and the propagation time in devices is $t = 2\pi R / c_0$. The total number of mesh elements is $N_V = 18602$; the average length of mesh edges is $\overline{\Delta}_\text{mesh} = 0.05 R$ and the step size takes $\Delta t = 10^{-5}~R/c_0$. The location of light source is $\vect{r}_\text{source} = (0, -0.01, 1.01)^\text{T}$ around the north pole of the surface $\mathcal{M}_\text{octa}$. The initial wave vectors are $\vect{k}_0^\text{A} = ( -0.58, 0.82, 0)^\text{T}$ and $\vect{k}_0^\text{B} = (-0.99, 0.12, 0)^\text{T}$. All the vectors are given in terms of the global Cartesian coordinates.}
    \label{fig:figure04}
\end{figure}

By considering the functional medium of \eqref{eq:Mobius-transformation-medium} in the virtual space, one can trace the light rays in the focal control device on the curved surface $\mathcal{M}_\text{octa}$ for focal control devices. The results of ray tracing in \figref{fig:figure04}{(a)} and \figref{fig:figure04}{(b)} validate the performance of the focal control device that light rays emitted from the designated point source $\vect{x}^i_\text{source}$ will pass through the expected focus point $\vect{x}^i_\text{focal}$. In comparison with points $\vect{Q}_1$ and $\vect{Q}_2$ in virtual space, the spatial region between source and focal points is shortened, where the higher refractive index distribution  $n^\text{(a)}_\text{max} = 2.75$ in \figref{fig:figure04}{(a)} and $n^\text{(b)}_\text{max} = 1.82$ in \figref{fig:figure04}{(b)} can compensate the optical path, so that the light trajectories A and B with different geodesic lengths can maintain the same optical path length. The minimum of refractive index is $n^\text{(a)}_\text{min} = 0.16$ and $n^\text{(b)}_\text{min} = 0.26$, both distributed on the bumps of the curved surface $\mathcal{M}_\text{octa}$. In addition, the trajectories A and B are both closed, which are consistent with the propagation behavior on the spherical surface $\mathcal{Q}$ in the virtual space. The source and focal points on $\mathcal{M}_\text{octa}$ correspond to points $\vect{M}_1$ and $\vect{M}_2$ marked in \figref{fig:figure01}{(a)} that are mapped from points $\vect{Q}_1$ and $\vect{Q}_2$ symmetrical about sphere center in virtual space $\mathcal{Q}$ through multiple mapping illustrated by \figref{fig:figure01}. 

Facing longer propagation distances and surfaces with greater intrinsic curvature, the emulation of the non-Euclidean metric by transformation medium $n_\text{trans}(F_{\mathcal{S}})$ will be damaged, and the optical devices may fail. Finer meshes can help emulate the curved metric better and improve the accuracy of the mesh-based ray tracing method. In addition, smaller step size $\Delta t$ can also bring better accuracy to ray tracing. By adjusting the mesh and step size in account of the intrinsic curvature of the surface, propagation time, and computing resources, a satisfying balance can be achieved between computing speed and precision.

\section{Conclusions}
In summary, we have proposed a general method for light control on a curved surface and a mesh-based ray-tracing algorithm on a surface filled with inhomogeneous medium. Within the framework of conformal transformation optics, the isotropic transformation medium can be induced from the change of singular values of the metric tensor during quasi-conformal mapping, and renders the propagation behavior on a curved surface identical to that in virtual space. A focal control device is proposed to designate the focus position of light rays emitted from a fixed source by repositioning a pair of points through the M\"{o}bius transformation as well as stereographic and inverse stereographic projection. We have unified the metric and medium as the optical metric and derived Hamilton's equations on a curved surface with medium. By solving the equations in the local coordinate system of mesh elements and illuminating the refraction caused by simplex non-planarity and medium discontinuity, the performance of the AOIs and focal control device on a rounded octahedral surface has been validated. We expect that the proposed light control method and ray-tracing algorithm will promote the research of electromagnetism with non-Euclidean geometry, transformation cosmology, and radar creeping wave.

\appendix
\renewcommand{\thesection}{\Alph{section}} 
\setcounter{equation}{0}
\renewcommand{\theequation}{A\arabic{equation}}
\section*{Appendix}
\subsection{Matrix form of metric tensor and coordinate transformations} \label{app:matrix-form-transformation}
The matrix forms of coordinate transformations \eqref{eq:uv-xyz-tensor} between the local coordinate system $\{u, v\}$ and global coordinate system $\{x, y, z\}$ are respectively
\begin{subequations}
    \begin{align}
        \begin{pmatrix}
        x \\
        y \\
        z 
    \end{pmatrix} &= 
    \begin{pmatrix}
        x_1 \\
        y_1 \\
        z_1 
    \end{pmatrix} + 
    \begin{pmatrix}
        x_2 - x_1 & x_3 - x_1 \\
        y_2 - y_1 & y_3 - y_1 \\
        z_2 - z_1 & z_3 - z_1
    \end{pmatrix}
    \begin{pmatrix}
        u \\
        v 
    \end{pmatrix} =
    \vect{p}_1 + [\vect{e}_u \ \vect{e}_v] \cdot
    \begin{pmatrix}
        u \\
        v 
    \end{pmatrix}, \\
    \begin{pmatrix}
        u \\
        v
    \end{pmatrix} &= 
    \begin{pmatrix}
        0 & 1 & 0 \\
        0 & 0 & 1
    \end{pmatrix}
    \begin{pmatrix}
        x_1 & x_2 & x_3 \\
        y_1 & y_2 & y_3 \\
        z_1 & z_2 & z_3
    \end{pmatrix}^{-1}
    \begin{pmatrix}
        x \\
        y \\
        z
    \end{pmatrix};
    \end{align}
\end{subequations}
and the Jacobi matrices $\Lambda^i_\alpha = {\partial x^i}/{\partial x^\alpha}$ and $J^\alpha_i = {\partial x^\alpha}/{\partial x^i}$ in \eqref{eq:uv-xyz-tensor} can be expressed as
\begin{subequations}
    \begin{align}
        \vect{\Lambda}_{uv2xyz} &=
    \begin{pmatrix}
        x_2 - x_1 & x_3 - x_1 \\
        y_2 - y_1 & y_3 - y_1 \\
        z_2 - z_1 & z_3 - z_1
    \end{pmatrix}, \label{eq:matrix-form-lambda} \\
    \vect{J}_{xyz2uv} &=  \begin{pmatrix}
        0 & 1 & 0 \\
        0 & 0 & 1
    \end{pmatrix}
    \begin{pmatrix}
        x_1 & x_2 & x_3 \\
        y_1 & y_2 & y_3 \\
        z_1 & z_2 & z_3
    \end{pmatrix}^{-1}. \label{eq:matrix-form-J}
    \end{align}
\end{subequations}
Now, the matrix form of the metric tensor $g_{\alpha\beta} = \vect{e}_\alpha \cdot \vect{e}_\beta$ reads
\begin{equation} \label{eq:ch04:local-coordinate-metric-matrix}
    \vect{g} = \begin{pmatrix}
        \vect{e}_u \cdot \vect{e}_u & \vect{e}_u \cdot \vect{e}_v \\
        \vect{e}_v \cdot \vect{e}_u & \vect{e}_v \cdot \vect{e}_v
    \end{pmatrix},
\end{equation}
where $\vect{e}_u = (x_2 - x_1, y_2 - y_1, z_2 - z_1)^\text{T}$ and $\vect{e}_v = (x_3 - x_1, y_3 - y_1, z_3 - z_1)^\text{T}$ are the basis vectors for the local coordinates $u$ and $v$, respectively. In addition, the matrix forms of transformations \eqref{eq:wave-vector-transformation} between the local coordinates and the global Cartesian coordinates for the wave vector are
\begin{subequations} \label{eq:kuv-and-kxyz}
\begin{align} 
    \vect{k}_\alpha &= (\vect{k}^i)^\text{T} \vect{J}_{xyz2uv}^\text{T} \vect{g} = 
    \begin{pmatrix}
        k_u & k_v
    \end{pmatrix}  \label{eq:kuv-and-kxyz-1} \\
    \vect{k}^i &= \vect{\Lambda}_{uv2xyz} \vect{g}^{-1} (\vect{k}_\alpha)^\text{T} =
    \begin{pmatrix}
        k_x & k_y & k_z
    \end{pmatrix}^\text{T}  \label{eq:kuv-and-kxyz-2}.
\end{align}
\end{subequations}

\subsection{Pseudo-code for ray tracing on a meshed surface} \label{app:pseudocode-ray-tracing}
\begin{algorithm}[H]
    \caption{Ray-tracing algorithm on a meshed surface filled with medium}\label{alg:ray-tracing-meshed}
    \begin{algorithmic}[1]
        \Procedure{Ray\_Tracing}{$u_0, v_0, k_{u0}, k_{v0}, \vect{V}, \vect{F}, F_\text{now}, c_0, n, \Delta t, T$} \\
        \Comment{$u_0, v_0$: initial location in the local coordinate system} \\
        \Comment{$k_{u0}, k_{v0}$: initial phase gradient} \\
        \Comment{$\vect{V}$: $N_V \times 3$ vertex array} \\
        \Comment{$\vect{F}$: $N_F \times 3$ mesh array} \\
        \Comment{$F_\text{now}$: index of initial mesh element} \\
        \Comment{$c_0$: light velocity in free space} \\
        \Comment{$n$: refractive index distribution} \\
        \Comment{$\Delta t$: solution step size} \\
        \Comment{$T$: maximum propagation time}
        \State $F_\text{next} \gets F_\text{now}$ 
        \Comment{Initialize the index of next mesh element}
        \State $t_\text{M} \gets 0$ 
        \Comment{Initialize the propagation time}
        \While{$t_\text{M} < T$} 
        \State $i \gets i+1$
        \Comment{Update iteration counting}
        \State $F_\text{last} \gets F_\text{now}$
        \Comment{Record index of last mesh element}
        \State $F_\text{now} \gets F_\text{next}$
        \Comment{Update index of next mesh element}
        \State $\vect{F}_\text{tri} \gets \vect{F}(F_\text{now}, :)$
        \Comment{Extract indices of vertices}
        \State $\vect{P} \gets \vect{V}(\vect{F}_\text{tri}, :)$
        \Comment{Extract coordinates of vertices}
        \If{$i > 1$}
            \State $[u_0, v_0, k_{u0}, k_{v0}] \gets $
            Refraction$[\vect{P}, \vect{V}, \vect{F}_\text{tri}, \vect{x}_{\overline{m}}, \vect{v}_\text{b},$ 
            \State $\qquad \qquad \qquad p_\text{near}, p_\text{far}, n(F_\text{last}), n(F_\text{now}), \beta, \vect{k}_m]$ \\
            \Comment{Calculate refraction \eqref{eq:rodrigues-rotation-simple}; update initial condition}
        \EndIf
        \State $[\vect{g}, Func] \gets $Metric$(P, n, c_0)$ \\
        \Comment{Obtain metric tensor and form of equations \eqref{eq:hamilton-isotropic-medium}}
        \State $[t_m, k_u, k_v, u, v] \gets$ Runge\_Kutta\_Solve$(u_0, v_0, k_{u0}, k_{v0},$ 
        \State $\qquad \qquad \qquad \qquad \qquad \qquad \qquad \qquad \quad \Delta t, T, \vect{g}, Func)$ \\
        \Comment{Solve the Hamilton's equation \eqref{eq:hamilton-isotropic-medium}}
        \State $\vect{x}_m \gets$ Global\_Coordinate$(P, u, v)$ \\
        \Comment{Obtain the global coordinates \eqref{eq:uv-2-xyz-tensor} of last point}
        \State $\vect{k}_m \gets$  Global\_Vector$(P, k_u, k_v, \vect{g})$ \\
        \Comment{Obtain the wave vector \eqref{eq:wave-vector-transformation-2} at last point}
        \State $[F_\text{next}, p_\text{near}, p_\text{far}] \gets$  Adjacent\_Triangle$(u, v, F_\text{now}, \vect{F}_\text{tri}, \vect{F})$ \\
        \Comment{Query next mesh element and boundary vertices}
        \State $[\vect{x}_{\overline{m}}, \vect{v}_\text{b}, \beta] \gets$  Boundary$[\vect{x}_m, \vect{V}(p_\text{near}, :), \vect{V}(p_\text{far}, :), \vect{k}_m]$ \\
        \Comment{Calculate boundary point \eqref{eq:location-boundary-point} and angle}
        \State $t_\text{M} \gets t_\text{M} + t_m$
        \Comment{Update propagation time}
        \EndWhile\label{euclidendwhile}
        \State \textbf{return} $u,v$ 
        \Comment{Output coordinates of light trajectory}
        \EndProcedure 
    \end{algorithmic}
\end{algorithm}

\subsection{Design of the rounded octahedral surface} \label{app:rounded-octahedral-surface}
The geometric model considered in \figref{fig:figure03} and \figref{fig:figure04} can be obtained by superimposing bumps or dents on the base surface of a sphere of radius $R$. Assume that the height of the bump or dent is $h_1$, the radius of the bottom is $R_1$, the center of the bottom is located at the point ${\vect{Q}} = (x_{{\vect{Q}}}, y_{{\vect{Q}}}, z_{{\vect{Q}}})^\text{T}$ on the sphere and the shape is represented by $z_\text{super}(x,y)$. If the bump or dent is superimposed on the upper hemisphere, the new surface above the $xy$ plane reads $z_\text{surf}(x,y) = z_\text{sphere}(x,y) + z_\text{super}(x,y)$, where $z_\text{sphere}(x, y) = \sqrt{R^2 - x^2 - y^2}$ describes the hemisphere $z > 0$; the part under the $xy$ plane keeps unchanged. Using the Gaussian distribution function to design the bump or dent, the function $z_\text{super}(x,y)$ can be expressed as $z_\text{super}(x, y) = h_1 \text{exp} \left\{ -\left[ (x-x_{\vect{Q}})^2 + (y-y_{\vect{Q}})^2 + (z_\text{surf}(x,y)-z_{\vect{Q}})^2 \right] / (2 R_1^2) \right\}$. Here, $h_1 > 0$ and $h_1 < 0$ mean bumps and dents, respectively. Set the bottom center of the bump or dent surface $\vect{Q} = (0, 0, R)^\text{T}$ at the north pole of the sphere, which can be regarded as a fixed superimposing position. The new surface after superimposition will rotate and repeat the superimposing procedure. The point $(x_0, y_0, z_0)^\text{T}$ on the surface after rotation will be $[x_1, y_1, z_1]^\text{T} = \vect{T}_x \cdot \vect{T}_y \cdot \vect{T}_z \cdot [x_0, y_0, z_0]^\text{T}$ with 
\begin{align}
    \vect{T} =& \begin{pmatrix}
            1 & 0 & 0 \\
            0 & \cos{\theta_x} & \sin{\theta_x} \\
            0 & -\sin{\theta_x} & \cos{\theta_x} \\
        \end{pmatrix} 
        \begin{pmatrix}
            \cos{\theta_y} & 0 & -\sin{\theta_y} \\
            0 & 1 & 0 \\
            \sin{\theta_y} & 0 & \cos{\theta_y} \\
        \end{pmatrix} \notag \\
        &\begin{pmatrix}
            \cos{\phi_z} & \sin{\phi_z} & 0 \\
            -\sin{\phi_z} & \cos{\phi_z} & 0 \\
            0 & 0 & 1 \\
        \end{pmatrix}
\end{align}
where $\theta_x$, $\theta_y$ and $\phi_z$ are the rotation angles around axis $x$, $y$ and $z$, respectively. In our calculation, the parameters for the rounded octahedral surface read: $R = 1$, $h_1 = 0.3$, $R_1 = 0.4$ and the rotation angles are $\theta_x = \pi/3$, $\phi_z = 0$; $\theta_x = \pi/3$, $\phi_z = 2\pi/3$; $\theta_x = \pi/3$, $\phi_z = 4\pi/3$; $\theta_x = 2\pi/3$, $\phi_z = \pi/3$; $\theta_x = 2\pi/3$, $\phi_z = \pi$; $\theta_x = 2\pi/3$, $\phi_z = 5\pi/3$.

\begin{backmatter}
\bmsection{Funding} National Natural Science Foundation of China (NSFC) 51977165.

\bmsection{Disclosures} The authors declare no conflicts of interest.

\bmsection{Data Availability Statement} All data needed to evaluate the conclusions are presented in the article. Raw data and the relevant code underlying the results presented in this paper are not publicly available at this time but may be obtained from the authors upon reasonable request.
\end{backmatter}

\bigskip

\balance
\bibliography{main}

\end{document}